\documentclass[conference]{IEEEtran}
\usepackage{cite}
\usepackage{float}
\usepackage[symbol]{footmisc}
\usepackage{amsmath,amssymb,amsfonts}
\DeclareMathOperator*{\argmin}{arg\,min}
\usepackage{graphicx}
\begin{document}

\title{Approaches to Simultaneously Solving Variational Quantum Eigensolver Problems}

\thanks{This research was supported in part by NSF Awards 2216923, 2117439, 2238734, 2217021 and 2311950}

\author{
    \IEEEauthorblockN{Adam Hutchings\IEEEauthorrefmark{1}\textsuperscript{\textsection}, Eric Yarnot\IEEEauthorrefmark{1}\textsuperscript{\textsection}, Xinpeng Li\IEEEauthorrefmark{1}, Qiang Guan\IEEEauthorrefmark{2}, Ning Xie\IEEEauthorrefmark{3}, Shuai Xu\IEEEauthorrefmark{1}, Vipin Chaudhary\IEEEauthorrefmark{1}}
    \IEEEauthorblockA{\IEEEauthorrefmark{1}Case Western Reserve University
    \\\{ash160, ery12, sxx214, xxl1337,  vxc204\}@case.edu}
    \IEEEauthorblockA{\IEEEauthorrefmark{2}Kent State University
    \\\ qguan@kent.edu}

    \IEEEauthorblockA{\IEEEauthorrefmark{3}Florida International University
    \\\ nxie@cis.fiu.edu}
}

\maketitle
\begingroup\renewcommand\thefootnote{\fnsymbol{footnote}}
\footnotetext[4]{Equal contribution}
\footnotetext[7]{This research was supported in part by NSF Awards 2216923, 2117439, 2238734, 2217021 and 2311950}
\endgroup


\begin{abstract}
The variational quantum eigensolver (VQE), a type of variational quantum algorithm, is a hybrid quantum-classical algorithm to find the lowest-energy eigenstate of a particular Hamiltonian. We investigate ways to optimize the VQE solving process on multiple instances of the same problem, by observing the process on one instance of the problem to inform initialization for other processes. We aim to take advantage of the VQE solution process to obtain useful information while disregarding information which we can predict to not be very useful. In particular, we find that the solution process produces lots of data with very little new information. Therefore, we can safely disregard much of this repetitive information with little effect on the outcome of the solution process.
\end{abstract}

\section{Introduction}

The variational quantum eigensolver (VQE) is a type of variational quantum algorithm (VQA), a class of hybrid quantum-classical algorithm \cite{vqa}. It has been used to solve several types of problems: quantum chemistry \cite{molecule2} \cite{qch} \cite{qchem}, traveling salesman \cite{traveling}, and solving the MaxCut problem \cite{maxcut}, among others. Given an arbitrary Hamiltonian $H,$ it seeks to find the eigenstate of this Hamiltonian $\psi_{min}$ with the lowest eigenvalue $\langle \psi_{min} | H | \psi_{min} \rangle.$ (It is of particular use in quantum chemistry because the configuration of a molecule can be expressed in such a way, and finding the lowest eigenvalue is akin to finding the ground state of a system. The VQE can therefore accurately predict molecular properties \cite{molecule}.) Given a circuit which produces an initial state $|\psi\rangle,$ the algorithm uses a classical optimizer to change the circuit such that the observed value $\langle \psi | H | \psi \rangle$ becomes smaller, gradually converging to a minimum. Noisy intermediate-scale quantum (NISQ) devices are likely to be the types of quantum devices practically available in the short term \cite{nisq}, but they have their limitations as many theoretical quantum algorithms require too many qubits to be used on such devices \cite{shor}.

However, there are issues with the current VQE strategy. For one, as with other variational quantum algorithms, there is a risk of the "barren plateau", where the optimization process halts prematurely as a result of being stuck in an area of the state space with low gradients \cite{vqeoverview}. Secondly, there may be many suboptimal local optima, which gradient descent can easily get stuck in. To mitigate these issues, choosing a state known to be close to the ground state is optimal, but in general such a method is not known. (Following terminology in \cite{xinpeng}, we will refer to states which the VQE evaluates as "points", as they may be thought of as vectors in some high-dimensional space.) We aim here to try to find a strategy for choosing good initial points.

Previous work in \cite{xinpeng} suggested that observing the solution process for one instance of a given problem (the seed problem) may be of some use for solving another (the target problem), by cataloguing points that the VQE algorithm considers while solving the seed problem and considering the data collected from these points while choosing an initial point for the target problem. In our particular case, we are given a seed problem with Hamiltonian $H_{seed}$ and target problems with Hamiltonians $H_1, H_2, \ldots.$ While running the VQE algorithm to find the minimum eigenstate of $H_{seed},$ we will consider a number of states $|\psi_1\rangle, |\psi_2\rangle, \ldots,$ and for a given target Hamiltonian $H_i$ we can then observe all of the $|\psi_k\rangle$ collected in the process of solving $H_{seed}.$ Previous work in \cite{xinpeng} has shown uses for this type of strategy in solving various problems in combinatorics.

We observe that as the VQE algorithm runs to completion on the seed problem, the points considered get much closer together in the state space. Therefore, we find that disregarding the large numbers of points which are extremely close to one another does not affect the results very much, speeding up the process of selecting an initial point in such simultaneous-solution strategies. Even when getting stuck in local optima in the state space, consideration of all points and consideration of only the first half tend to converge to the same local optima. This is an indicator that the two methods likely end up selecting the same starting point most of the time, confirming the hypothesis that the large numbers of close-together points evaluated near the end of the VQE process are not very useful and may easily be ignored.

\section{Approaches to Simultaneous Solving}

\subsection{The MaxCut Problem}

The VQE can be used on any problem, so long as it can be expressed as finding the minimum eigenstate of a Hamiltonian. Here we investigate the MaxCut problem \cite{maxcut} on a graph as a generic test algorithm to investigate the effectiveness of various parameter reuse strategies.

Given a graph $G = (V, E),$ the MaxCut problem aims to find a partition $V = V_1 \cup V_2$ of the set of vertices, such that the number of edges going between $V_1$ and $V_2$ is maximized. For the remainder of the discussion, say we have $k$ vertices.

To encode this problem into one which can be solved by a VQE, we will encode the vertices into a $k$-bit vector $s,$ whose entries $s_i$ are either one or zero. More specifically, we are looking to represent the problem such that $s_i$ is 0 if $v_i \in V_1,$ and 1 if $v_i \in V_2.$ Next, define a weight function $w_{ij} : \{1 \ldots k \}^2 \rightarrow \mathbb R^+$ which in this case is 0 if there is no edge connecting vertices $i$ and $j$ and is 1 if there is such an edge.

Then, given a vector $s,$ the number of edges going from $V_1$ to $V_2$ is
$$C(s) = \sum_{i < j} w_{ij}s_i(1-s_j).$$

The Ising Model then can take this and convert it into a Hamiltonian
$$H = \sum_{i < j} w_{ij}Z_iZ_j.$$

With this Hamiltonian, the MaxCut problem can now be solved by finding the minimum energy state using a VQE.

\subsection{Observing Points on Target Graphs}

When using a VQE to solve a problem, one issue that arises is the choice of selecting an initial point to begin optimization from. A poorly selected initial point can take many iterations to converge to an optimum, or become stuck on a barren plateau.

Work in \cite{xinpeng} proposed an approach to solving multiple MaxCut problems simultaneously. Because the VQE algorithm produces and evaluates many points before converging to a solution, when the VQE is being run on one (seed) graph, we can observe and evaluate these points on other (target) graphs. Then, when it comes time to solve said target graphs, we can consider the evaluations produced by all points from the seed graph when selecting an initial point. This may be helpful because we are solving a problem which is similar in some regards, so feasible solutions for one problem \emph{may} be feasible for the other, as shown for MaxCut instances on graphs in the QAOA paradigm \cite{randomtransfer} \cite{parametertransfer}. As a continuation of the methods in \cite{xinpeng}, we do not consider similarity properties between the problem instances, and see whether just the fact that they are the same type of problem might help.

However, we notice that in the process of convergence to a minimum in a MaxCut problem, the points considered in the state space become much closer together as the process of optimization goes on. In particular, for graphs we have collected data on, the points in the second half of the search are clustered very closely together. Presumably, these points are near a local minimum on the seed graph (where the VQE process was actually run), but they are unlikely to be near a local minimum on any target graph, exponentially so in the number of graph vertices.

Therefore, our second approach is to only consider the points in the first half of the solution process, because the second half will not give us much new information.

Additionally, in case observing seed problems does not help solving a target problem, we test a strategy which disregards other solutions entirely, by simply selecting a random initial point.

\begin{figure}[H]
    \centering
    \includegraphics[width=8cm]{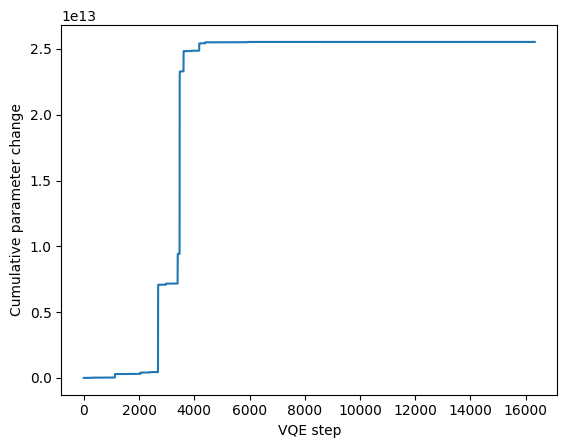}
    \caption{The sum total of how much parameters change between iterations, with some given metric, to illustrate how the VQE process tends to work on a real problem. Clearly, the parameters barely change at all in the later stages of the solution process, so the points are extremely close together in the state space.}
\end{figure}

\subsection{Three Approaches}

We will produce random graphs $G_1 \ldots G_k,$ of $n$ nodes each, where each edge has probability $\tilde p$ to be included in the graph. In the below discussion, we will use the following notation:

\begin{itemize}

\item
Given a point $p,$ $E_i(p)$ represents the observation value of $p$ on $G_i.$

\end{itemize}

We will test three approaches to finding an initial point for the VQE process on each graph:

\begin{enumerate}

\item
Given another random graph $G,$ we solve the MaxCut problem on $G,$ while observing the expectation values of the produced points $p_1 \ldots p_I$ on all $G_1 \ldots G_k.$ Then, for each $G_i,$ we choose as our initial point $p_{j_i},$ where
$$j_i = \argmin_{h \in \{1 \ldots l \}} |E_i(p_h)|.$$

\item
We do the same process, but only select the best points out of the first half (where convergence is not quite so slow). In other words, we choose $p_{j_i}$ where
$$j_i = \argmin_{h \in \{1 \ldots \frac l 2 \}} |E_i(p_h)|.$$

\item
The null hypothesis: we randomly select a point $p$ and use it on all $G_1 \ldots G_k.$

\end{enumerate}

For each method, we will analyze the number of iterations taken for the VQE to run in each instance, as well as the finishing energy, and average them.

\section{Experimental Setup}

We tested our method on $n = 5, 6, 7$ and $10$ node graphs. We generated $k = 9$ random graphs with edge probability $\tilde p = 1/2$ (except in the case of the $10$ node graphs, where we generated only $2$ in the interest of time), and found their associated Ising Hamiltonians using Qiskit. For each particular problem, the associated Hamiltonian is the observable we were trying to optimize. For the VQE process, the ansatz was Qiskit's TwoLocal circuit with linear entanglement. The number of repetitions of alternating rotation and entanglement layers, as the TwoLocal circuit contains, we varied from two to five, specifics of which are discussed in our results. (As a result, the number of tunable parameters varied -- for a ballpark, the case of 6-node graphs with two layers of repetition gave 18 parameters in the circuit.)

We then arbitrarily selected our last graph to be our seed one, and found its minimum eigenvalue using the Qiskit VQE algorithm evaluated using the Aer simulator. We used the Qiskit-provided Sequential Least Squares Programming (SLSQP) Optimizer for the VQE process (a test with an alternate optimizer, the SPSA optimizer, provided markedly worse results). 

The parameters at each step of the seed graph solution were saved. Then, for each of the 8 target graphs, we selected every 10th parameter set and calculated the expectation value for each set. This resulted in a list of parameter / energy pairs which we could select minimal values from for methods 1 and 2.

We took these values and used them as the initial points for the Qiskit VQE algorithm. Our random strategy was run by providing no initial point, so the algorithm will generate a random one within bounds. We then saved the data and averaged the results over the 8 target graphs.

We repeated this process 10 times and averaged all the results.

\section{Experimental Results}

After averaging our data, the following results were observed:

\begin{figure}[H]
    \centering
    \includegraphics[width=8cm]{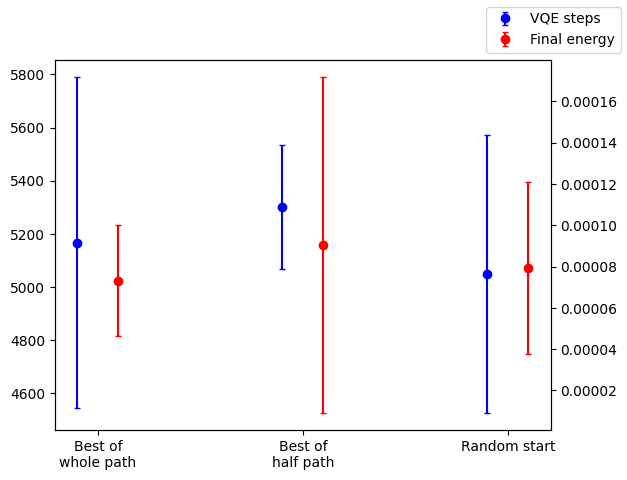}
    \caption{Results for our three strategies: considering all points in the seed graph, considering only half of the points, and considering none of the points. This is the run on 5-node graphs, with a TwoLocal circuit with five layers.}
\end{figure}

\begin{figure}[H]
    \centering
    \includegraphics[width=8cm]{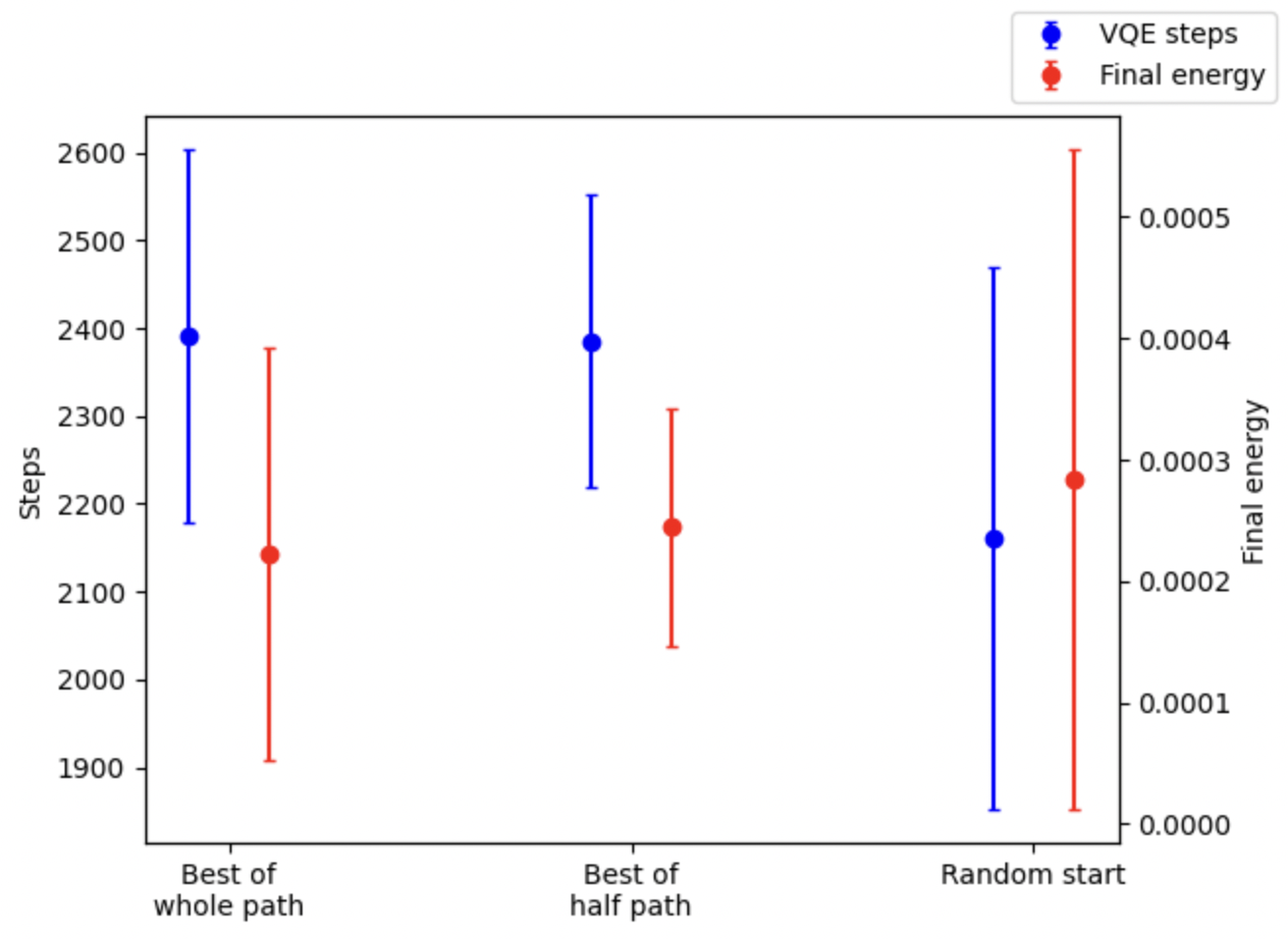}
    \caption{Results for a TwoLocal circuit with 2 layers, on 6-node graphs.}
\end{figure}

\begin{figure}[H]
    \centering
    \includegraphics[width=8cm]{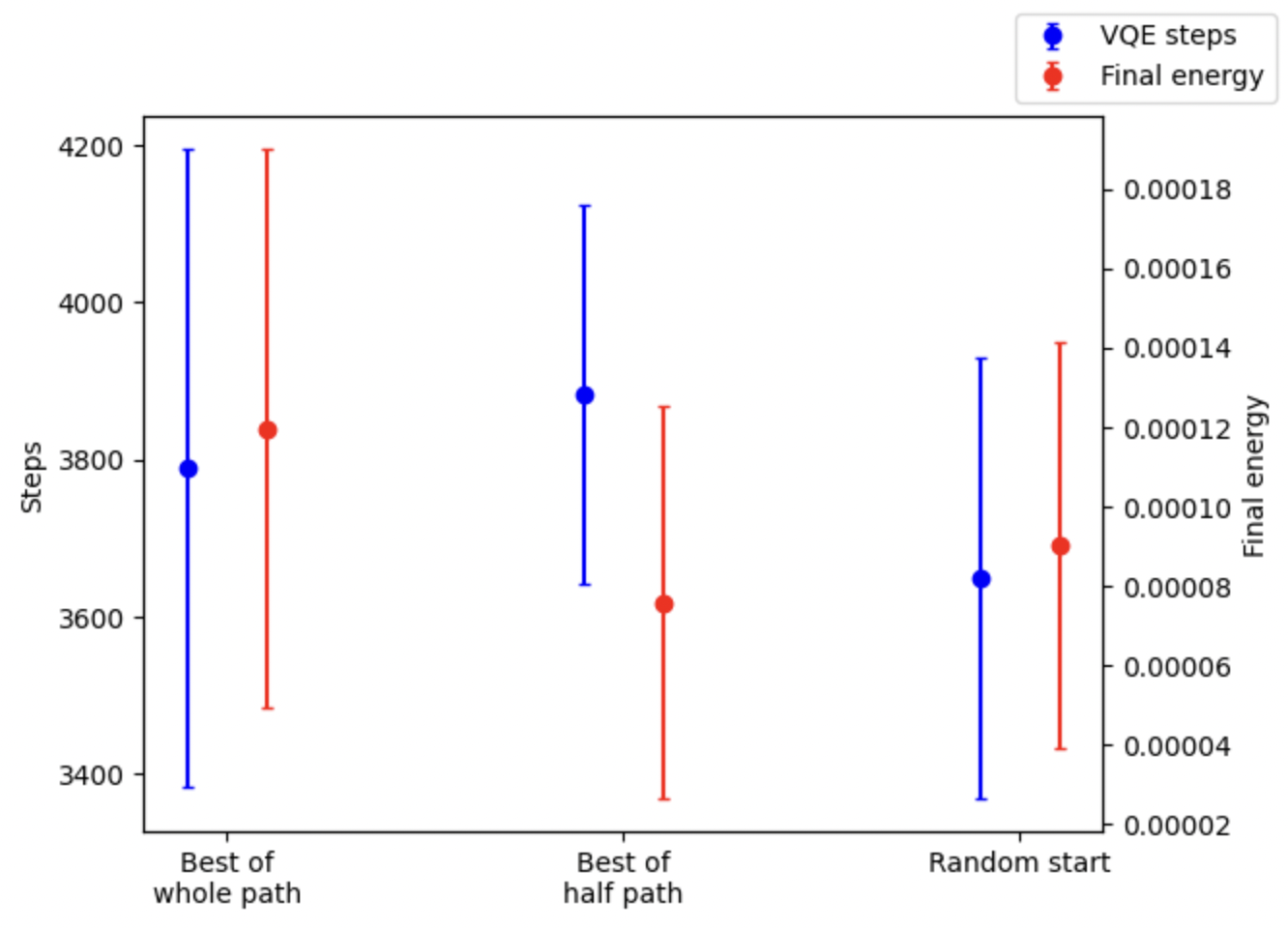}
    \caption{Results for a TwoLocal circuit with 3 layers, on 6-node graphs.}
\end{figure}

\begin{figure}[H]
    \centering
    \includegraphics[width=8cm]{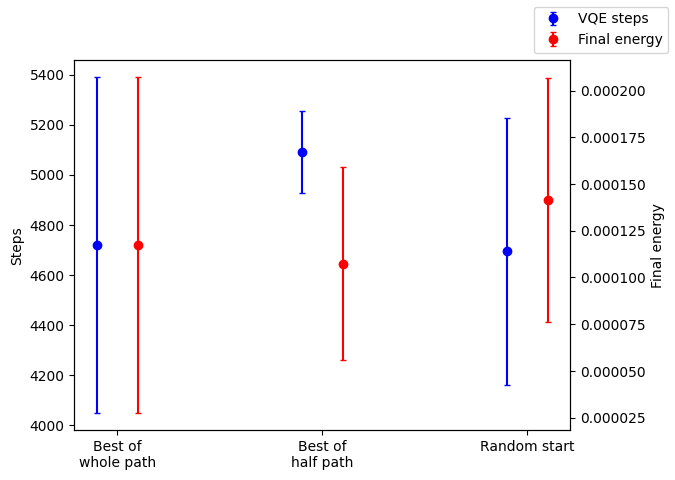}
    \caption{Results for a TwoLocal circuit with 3 layers, on 7-node graphs.}
\end{figure}

Two main results are apparent in these values. Firstly, choosing a random point performs about as well as selecting a point from observing of a seed problem, \emph{given our current selection strategy}. (Work in \cite{xinpeng} suggests other evaluation strategies that may be worth considering to achieve better results compared to chance.) Secondly, only observing half of the points to make a decision on where to start does not affect the result too much, which we can see by noting that the estimated ground-state energies for the first two strategies are quite similar. In fact, we observed that the first-half and full observation strategies tended to converge to the same solution. These patterns held for the $10$-node graphs as well. Further, in terms of the ground-state energies actually estimated on individual problem instances with the two methods, these varied significantly from graph to graph but were broadly similar between the methods. Together, this implies that neither strategy offered a great advantage over the other in solution quality. Given this, we find that selecting from only the first half of points may be a preferred method, as it takes less computation to arrive at a similar solution and also incurs less of a storage overhead.

While all methods perform about equally now, this is likely because of our very simple selection strategy (choosing the state with the lowest observed energy), which can be improved by considering other factors as in \cite{xinpeng}. The key result we see is that we may disregard many of the observed points from the seed graph solution process without issue, and combining this insight with prior work may result in more effective solution processes.

\section{Conclusion}

While our results are still preliminary and more data would be of great help, we confirm our expectations that disregarding many observed points from consideration does not affect our overall results too much. Because the points considered at the end of the VQE process on the seed graph are all very close together and \emph{a priori} quite unlikely to be near the global minimum on any target graph, simply ignoring them does not affect the target graph results by a lot.

However, our results also show that the methods both perform roughly as well as choosing an initialization point completely randomly, without consideration of the solution process for other graphs. This may be due to the fact that we are only considering the energy value of points we consider, without considering other factors such as gradients which may be helpful \cite{xinpeng}. In any case, though, for a strategy which involves considering the solution process on a seed problem to initialize the solution on a target problem, we find that many points do not need to be considered at all.

\end{document}